\documentclass{aa}

\usepackage[varg]{txfonts}
\usepackage[]{natbib}
\usepackage{graphicx}
 
\usepackage{dsfont}
\usepackage{esdiff}
\usepackage{amssymb}
\usepackage{wasysym}

\usepackage{siunitx}
\DeclareSIUnit{\year}{yr}
\DeclareSIUnit{\gauss}{G}
\DeclareSIUnit{\Jupiterradius}{\emph{R}_{\jupiter}}
\DeclareSIUnit{\Jupitermass}{\emph{M}_{\jupiter}}

\newcommand{\subs}[2]{{#1}_\mathrm{#2}}
\newcommand{\dint}{\mathrm d}

\bibpunct{(}{)}{;}{a}{}{,} % to follow the A&A style

\begin{document}

\title{Shallowness of circulation in hot Jupiters}
\subtitle{Advancing the Ohmic dissipation model}
\author{
H.~Knierim \inst{\ref{inst1}}
\and K.~Batygin \inst{\ref{inst2}}
\and B.~Bitsch \inst{\ref{inst1}}
}
\institute{
Max-Planck-Institut für Astronomie, Königstuhl 17, Heidelberg 69117, Germany \label{inst1}
\and
Division of Geological and Planetary Sciences
California Institute of Technology
Pasadena, CA 91125, USA
\label{inst2}
}
\date{Received 5 November 2021 / Accepted 19 January 2022}
\abstract{The inflated radii of giant short-period extrasolar planets collectively indicate that the interiors of hot Jupiters are heated by some anomalous energy dissipation mechanism. Although a variety of physical processes have been proposed to explain this heating, recent statistical evidence points to the confirmation of explicit predictions of the Ohmic dissipation theory, elevating this mechanism as the most promising candidate for resolving the radius inflation problem. In this work, we present an analytic model for the dissipation rate and derive a simple scaling law that links the magnitude of energy dissipation to the thickness of the atmospheric weather layer. From this relation, we find that the penetration depth influences the Ohmic dissipation rate by an order of magnitude. We further investigate the weather layer depth of hot Jupiters from the extent of their inflation and show that, depending on the magnetic field strength, hot Jupiter radii can be maintained even if the circulation layer is relatively shallow. Additionally, we explore the evolution of zonal wind velocities with equilibrium temperature by matching our analytic model to statistically expected dissipation rates. From this analysis, we deduce that the wind speed scales approximately as $1/\sqrt{T_\mathrm{eq}-T_0}$, where $T_0$ is a constant that equals $T_0 \sim \SI{1000}{\kelvin}-\SI{1800}{\kelvin}$ depending on planet-specific parameters (radius, mass, etc.). This work outlines inter-related constraints on the atmospheric flow and the magnetic field of hot Jupiters and provides a foundation for future work on the Ohmic heating mechanism.}
\maketitle
%%
%Section: Introduction
%
\section{Introduction}\label{sec:intro}
Generic models of giant planetary structures hold that the radii of evolved Jovian-class planets cannot significantly exceed the radius of Jupiter \citep{Stevenson_1982}. Accordingly, measurement of the substantially enhanced radius of HD 209458 b -- clocking in at $ \SI{1.39}{\Jupiterradius}$ \citep[$\SI{0.73}{\Jupitermass}$, $ \SI{1476.81}{\kelvin}$;][]{Stassun_2017} -- came as a genuine surprise \citep{Charbonneau_2000, Henry_2000}. Subsequent detections demonstrated that HD 209458 b is not anomalous; rather, inflated radii are a common attribute of the hot Jupiter (HJ) class of exoplanets \citep{Bodenheimer_2003, Laughlin_2011}. This begs the obvious question of what physical mechanism inflates the radii of HJs.\

Although a number of hypotheses have been proposed -- for example: tidal heating \citep[e.g.,][]{Bodenheimer_2001}, thermal tides \citep[e.g.,][]{Arras_2010}, advection of potential temperature \citep[e.g.,][]{Tremblin_2017}, or the mechanical greenhouse effect \citep[e.g.,][]{Youdin_2010} -- recent data indicate that explicit predictions of the Ohmic dissipation (OD) mechanism \citep[e.g.,][]{Batygin2010, Perna_2010b, Perna_2010, Wu_2012, Spiegel_2013, Ginzburg_2016} are consistent with the modern data set. Specifically, the functional form of the inflation profile, and its (Gaussian) dependence on the planetary equilibrium temperature, is reflected in the data \citep{Thorngren_2018, Sarkis:2021tr}.\footnote{The other inflation models scale like power laws, meaning that their efficiency keeps increasing with equilibrium temperature.}
In light of this, in this work we aim to advance the theoretical understanding of the OD mechanism and pose the simple question of how Ohmic heating depends on the depth to which large-scale atmospheric circulation penetrates. Additionally, we examine if the empirical inflation profile derived from the data themselves can constrain the shallowness of circulation layers on these highly irradiated planets.

General circulation models (GCMs) based on “primitive” equations of hydrodynamics routinely find winds in the $ \si{\km \per \s}$ range that penetrate to depths on the order of 1 to 10 bars \citep[e.g.,][]{Showman_2015, Rauscher_2013}. \cite{Menou_2020}, however, argues that this depth can actually be significantly greater. Analytic models based on force-balance considerations \citep[][]{Menou2012, Ginzburg_2016} or models that regard the HJ atmosphere as a heat engine \citep[][]{Koll_2018} have predicted similarly fast winds, with magnetic drag dominating at high equilibrium temperatures.
On the contrary, anelastic magnetohydrodynamic (MHD) simulations find rather shallow circulation layers with slow wind velocities in general \citep[e.g.,][]{Rogers_2014, Rogers_2014b, Rogers_2017}.

Given that each of these models is subject to its own strengths and limitations, an alternative constraint on HJ circulation patterns is desirable. Accordingly, by extending the Ohmic heating model to include varying zonal wind geometries, we demonstrate the existence of a simple scaling law that allows us to predict the wind depth of HJs from the extent of their inflation under the assumption that Ohmic heating constitutes the dominant interior energy dissipation mechanism.

Another point of considerable interest is the evolution of wind patterns with equilibrium temperature. Force balance arguments suggest strong magnetic damping of equatorial winds at high equilibrium temperatures, $\subs{T}{eq}$ \citep{Menou2012}. In contrast, MHD simulations show flow reversal (i.e., eastward jets become westward) with increasing $\subs{T}{eq}$ \citep{ Batygin2013, Heng_2014,Rogers_2014}. Bypassing the question of wind direction, in this work we explore the mean equatorial wind velocity from energetic grounds using statistical evidence of the heating efficiency \citep{Thorngren_2018}.

The remainder of the manuscript is organized as follows. In Sect. \ref{sec:theory} we outline the HJ and OD model. We present the resulting circulation profiles in Sect. \ref{sec:results}. We discuss our results and conclude in Sect. \ref{sec:discussion}.
%
%Section: Theory
%
\section{Theory}\label{sec:theory}
\subsection{Conductivity profile}
The machinery of the OD mechanism is underlined by the fact that -- owing to the thermal ionization of alkali metals that are present in trace abundances -- HJ atmospheres are rendered weakly conductive. Accordingly, the first step in our analysis is to delineate the conductivity profile. We divided the HJ into radiative and convective layers, with the radiative-convective boundary (RCB) located at $ \SI{100}{bars}$.\footnote{A more complete approach would solve the Schwarzschild criterion to obtain the RCB pressure. Because our results depend only weakly on the exact value, we approximate it to be $ \SI{100}{bars}$ for consistency with \cite{Batygin2010}. Recent statistical studies \citep{Thorngren2019, Sarkis:2021tr} arrive at considerably shallower RCB values, which slightly expands our weather layer size (see Appendix \ref{sec:RCB_depth}).}

In practice, the functional form of the conductivity profile is only important in the vicinity of the RCB. In the radiative layer, we approximated the temperature as isothermal by averaging over the plane-parallel static gray atmosphere \citep{Guillot2010}:
\begin{align}
\begin{split}
T^4 &= \frac{3\subs{T}{int}^4}{4} \left(\frac{2}{3}+ \tau \right)\\
&+ \frac{3\subs{T}{eq}^4}{4}\bigg\{\frac{2}{3} + \frac{2}{3\zeta}\left[1 + \left( \frac{\zeta \tau}{2} - 1\right) e^{-\zeta \tau} \right]\\
&+\frac{2 \zeta}{3}\left(1-\frac{\tau^2}{2}\right) E_2(\zeta \tau)\bigg\},
\end{split}
\end{align}
where $T$ is the temperature in the atmosphere, $ \subs{T}{int}$ the intrinsic temperature, $ \subs{T}{eq}$ the equilibrium temperature of the HJ, $ \tau$ the optical depth, $\zeta := \frac{\subs{\kappa}{v}}{\subs{\kappa}{th}}$ the opacity ratio, and $E_n(z) := \int_1^\infty t^{-n} e^{-zt} \dint t$ the exponential integral function. We adopted \citet{Guillot2010} values for the visible and thermal opacity of HD 209458 b: $ \subs{\kappa}{v} = \SI{0.004}{\cm \squared \per \g}$ and $ \subs{\kappa}{th} = \SI{0.01}{\cm \squared \per \g}$, respectively.

Below the RCB, we assumed a polytropic relation $P \propto \rho^\gamma$, where $P$ is the pressure and $ \gamma$ is the polytropic index. We obtained $\gamma$ by fitting along an adiabat from the hydrogen-helium equation of state of \cite{Chabrier_2021}. Because the conductivity rises rapidly in the interior, the contribution of the deep interior to the heating is small. Thus, our polytropic fit suffices for our purposes. To this end, we note that anomalous heating need not be deposited at the center of the convective envelope to sustain the planetary radius in an inflated state. That is to say, energy deposited sufficiently deep in the envelope, a radius that translates to a pressure level on the order of $ \SI{10}{\kilo \bar}$~\citep[][and the references therein]{Batygin2011,Spiegel_2013, Ginzburg_2016, Komacek2017}, is sufficient. Reinflating a HJ from a cold state, however, requires deeper heat deposition \citep[e.g.,][]{Thorngren2021}. Generally, concerns of Ohmic reinflation of close-in giant planets \citep{Hartman_2016, Grunblatt_2017} -- while fascinating -- are beyond the scope of this work.

The ionization fraction of alkali metals is determined by Saha's equation:
\begin{align}
\frac{n_j^+ n_\mathrm{e}}{n_j - n_j^+} = \left(\frac{m_\mathrm{e} k T}{2 \pi \hbar^2}\right)^{3/2} \exp \left( -\frac{I_j}{k T}\right),
\end{align}
where the subscript $j$ denotes the respective alkali metal, $n_j^+$ its positively ionized number density, $I_j$ its ionization potential, and $n_j$ its total number density and $k$ is the Boltzmann constant, $ \subs{m}{e}$ the electron mass, $n_\mathrm{e}$ the electron number density, and $\hbar$ the reduced Planck constant. We can solve this equation in the weakly ionization limit ($n_j^+ \ll n_j$) to obtain the electron number density. Throughout this study, we use the proto-solar abundances of \cite{Lodders2003}, $Z_\odot=0.0149$, to obtain $n_j$.

Assuming a classical free electron gas, the conductivity, $ \sigma$, in the isothermal layer simplifies to an exponential function \citep{Batygin2010}:
\begin{align}\label{eq:iso_conductivity}
\sigma (r) =  \subs{ \sigma}{RCB} \exp \left( \frac{r- \subs{r}{RCB}}{2 H}\right),
\end{align}
where the subscript RCB denotes the respective quantity value at the RCB, $r$ the radial distance, and $H$ the pressure scale height.

The adiabatic conductivity expression is more complex. While a semi-analytic profile can in principle be readily derived, for simplicity we approximate this layer with an exponential profile as well: $\sigma =  \subs{ \sigma}{RCB} \exp \left[ (\subs{r}{RCB}-r)/\subs{H}{fit}\right]$,
where $\subs{H}{fit}$ is the free parameter used to fit the profile to the adiabatic expression.

\begin{figure}
\centering
        \resizebox{\hsize}{!}{
                \includegraphics{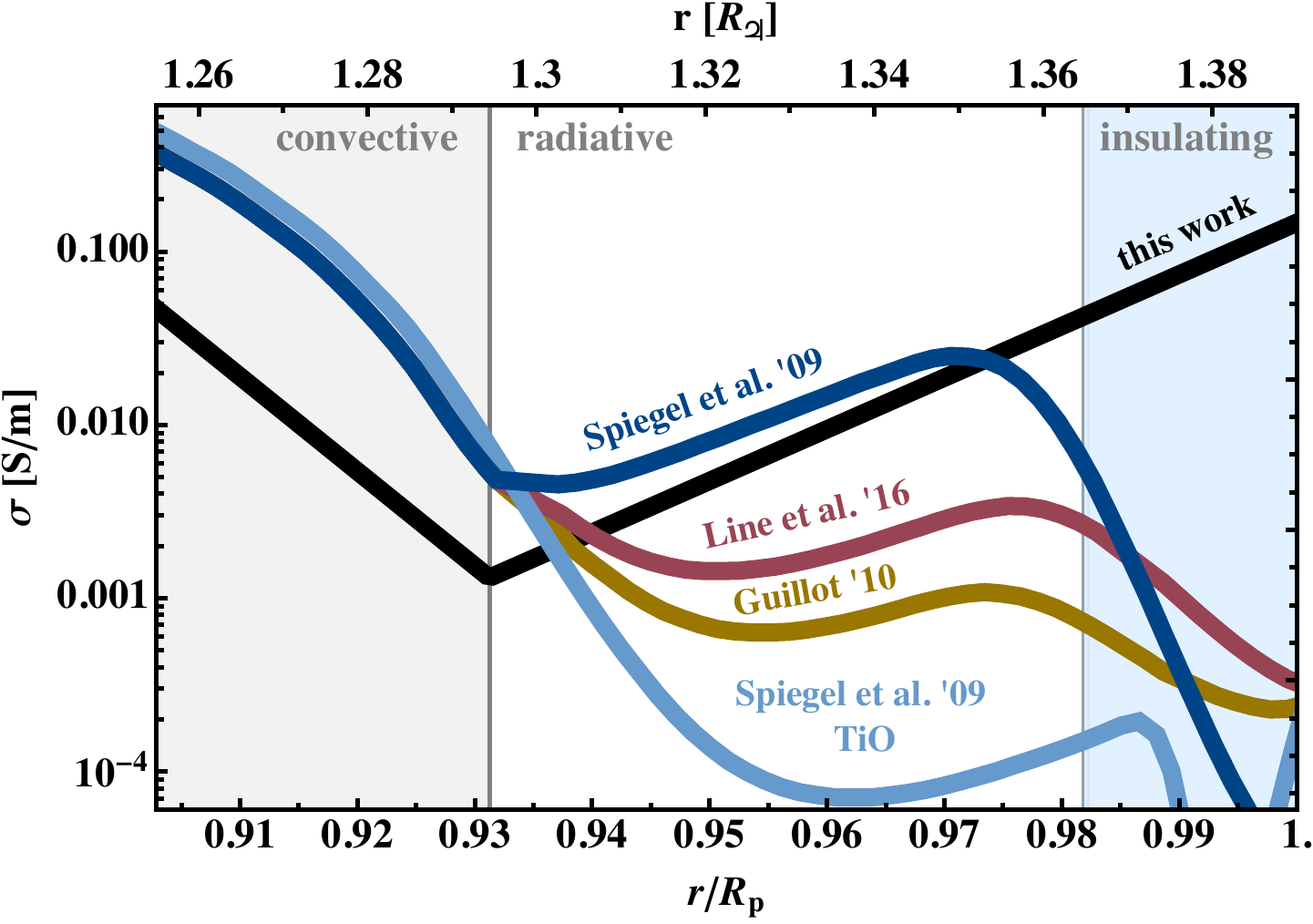}
        }
        \caption{Conductivity profile of this work (black) in comparison with that of \cite{Kumar_2021} for HD 209458 b. The references above the lines indicate the atmosphere model used by \cite{Kumar_2021} to compute the respective conductivity profile, namely: \cite{Guillot2010}, \cite{Line_2016}, and \cite{Spiegel:2009vi} with and without TiO in the atmosphere.}
        \label{fig:conductivity}
\end{figure}

Recently, \cite{Kumar_2021} undertook a sophisticated calculation of $\sigma$. We did not reproduce their treatment but compare their results with our simple estimate in Fig. \ref{fig:conductivity}. Crucially, all conductivity profiles show a drop close to $R \sim \SI{1.365}{\Jupiterradius} = 0.98~R_\mathrm{p}$, and we modeled this with an insulating boundary condition at $r = R$.
%
%Subsection:Velocity Profile
%
\subsection{Velocity profile}
Most HJs orbit their host star on circular orbits \citep{Bonomo_2017}. Tidal de-spinning will lock those planets, creating a constantly irradiated {dayside} and a {nightside} that faces away from the host star \citep{Hut_1981}. This unbalance in heating leads to a pressure gradient, which balances with the planet's rotation and drag to create strong winds on HJs \citep[for a review, see][]{Showman_2012}. For HD 209458 b, for example, \citet{Snellen2010} observed wind velocities of around $\SI{2\pm1}{\km \per \s}$ in the upper atmosphere.

Various GCMs \citep[e.g.,][]{Heng_2011c, Parmentier_2013, Showman_2015, Menou_2019}, analytical theory \citep[e.g.,][]{Showman_2010, Showman_2011}, and observations of hot-spot shifts \citep[e.g.,][]{Knutson_2007} indicate that these winds manifest as a single dominant eastward jet. In the presence of $10$ -- $100~\si{\gauss}$ magnetic fields \citep{Yadav2017, Cauley2019}, significant induction of electrical currents is unavoidable. While the conductivity of the atmosphere is insufficient for the resulting Lorentz force to fully suppress the circulation \citep{Batygin2013}, conductivity rises rapidly in the convective interior, ensuring the dominance of the magnetic drag \citep[e.g.,][]{Liu2008, Rogers_2014b}. As a consequence, zonal winds will be dissipated completely at or below the RCB (velocity $\vec{v}=0$ for $r \leq \subs{r}{RCB}$).

\begin{figure}
\centering
        \resizebox{\hsize}{!}{
                \includegraphics{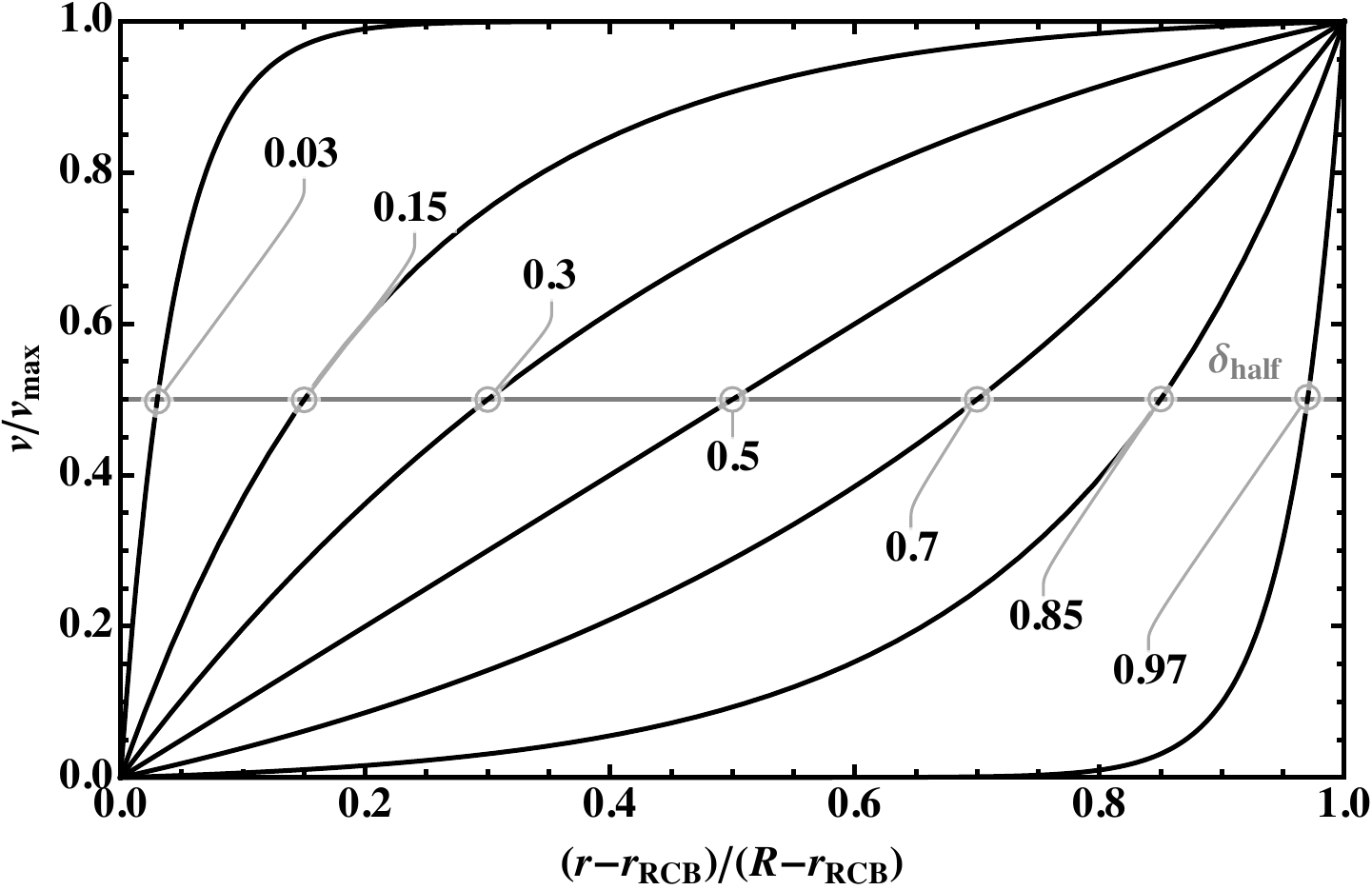}
        }
        \caption{Zonal wind velocity at the equator (Eq. \ref{eq:velocity}) as a function of radius for five different relative half-velocity radii, $ \subs{\delta}{half}$.}
        \label{fig:velocity}
\end{figure}

This led us to parametrize the velocity field in the radiative layer by
\begin{align}\label{eq:velocity}
\vec{v}(r,\theta)= \subs{v}{max}\cdot\left(\frac{e^{ \xi r}-e^{\xi \subs{r}{RCB}}}{e^{ \xi R}-e^{\xi \subs{r}{RCB}}}\right)\sin \theta
 \cdot \hat \phi,
\end{align}
where $\subs{v}{max}$ is the maximum velocity at the equator, $R$ the OD radius, $\theta$ the polar angle, and $\hat \phi$ the basis vector in azimuthal direction. The $\xi$ is a free parameter of dimension 1/length that controls the radial form of the zonal jet by damping or enhancing the penetration depth.
To simplify the comparison between various velocity profiles\footnote{In principle, a number of velocity profiles would fit our requirements. For example, the power law $ \left(\frac{r- \subs{r}{RCB}}{R- \subs{r}{RCB}}\right)^\xi$ also works. However, the derivative of this profile diverges for $0<\xi<1$, introducing unnecessary complications into the analytical treatment.} and give a physical intuition, we introduced the (dimensionless) relative half-velocity radius, $ \subs{\delta}{half}$, which is the relative distance between the RCB and the OD radius where $v$ decreased by $1/2$ (Fig. \ref{fig:velocity}).
If $\subs{\delta}{half}=0.5$, the velocity decreases linearly toward the RCB. If $0.5<\subs{\delta}{half}<1$, the velocity drops off more quickly and the zonal jets are confined to a smaller region in the upper atmosphere. Lastly, if $ 0<\subs{\delta}{half}<0.5$, the velocity decreases more slowly and the zonal jet is more extended.
%Subsecton Ohmic Dissipation:
\subsection{Ohmic dissipation rate}\label{sec:OD_theory}
Since the introduction of the OD model \citep{Batygin2010, Batygin2011}, a number of authors have explored various dependences of this heating mechanism \citep[e.g.,][]{Wu_2012, Spiegel_2013, Ginzburg_2016}, and the basic theoretical framework of OD is well documented in the literature. Hence, we will skip the derivation here and only state the equations that we are solving.

To obtain the OD rate, $\mathds{P}$, we need to compute\begin{align}\label{eq:Ohmic dissipation rate}
\mathds{P} = \int \frac{ \subs{\vec j}{ind}^2}{\sigma} \dint V,
\end{align}
where $\dint V$ is the volume element and $ \subs{\vec j}{ind}$ is the induced current density.
We can obtain $\subs{\vec j}{ind}$ from Ohm's law:
\begin{align} \label{eq:j_ind}
\subs{\vec j}{ind}=\sigma (\vec v \times \vec B-\nabla \Phi),
\end{align}
where $\vec B$ is the HJ's magnetic field, which we approximate by a dipole, and $\Phi$, the electric potential, is obtained by solving a variation of Poisson's equation that comes from the continuity of the current density:
\begin{align}\label{eq:electric_potential}
\nabla \cdot \sigma \nabla \Phi = \nabla \cdot \sigma \left(\vec{v} \times \vec{B}\right).
\end{align}
The radius of an inflated gas giant is predominately determined by its interior entropy \citep{Zapolsky_1969}. Heat deposited into the radiative layer radiates away before influencing the adiabat noticeably \citep{Spiegel_2013}. Consequently, Eq. \ref{eq:Ohmic dissipation rate} is restricted to evaluation in the convective zone when determining the radius inflation.
%Section Results
\section{Results}\label{sec:results}
\subsection{Weather layer depth}
\begin{figure}
\centering
        \resizebox{\hsize}{!}{
                \includegraphics{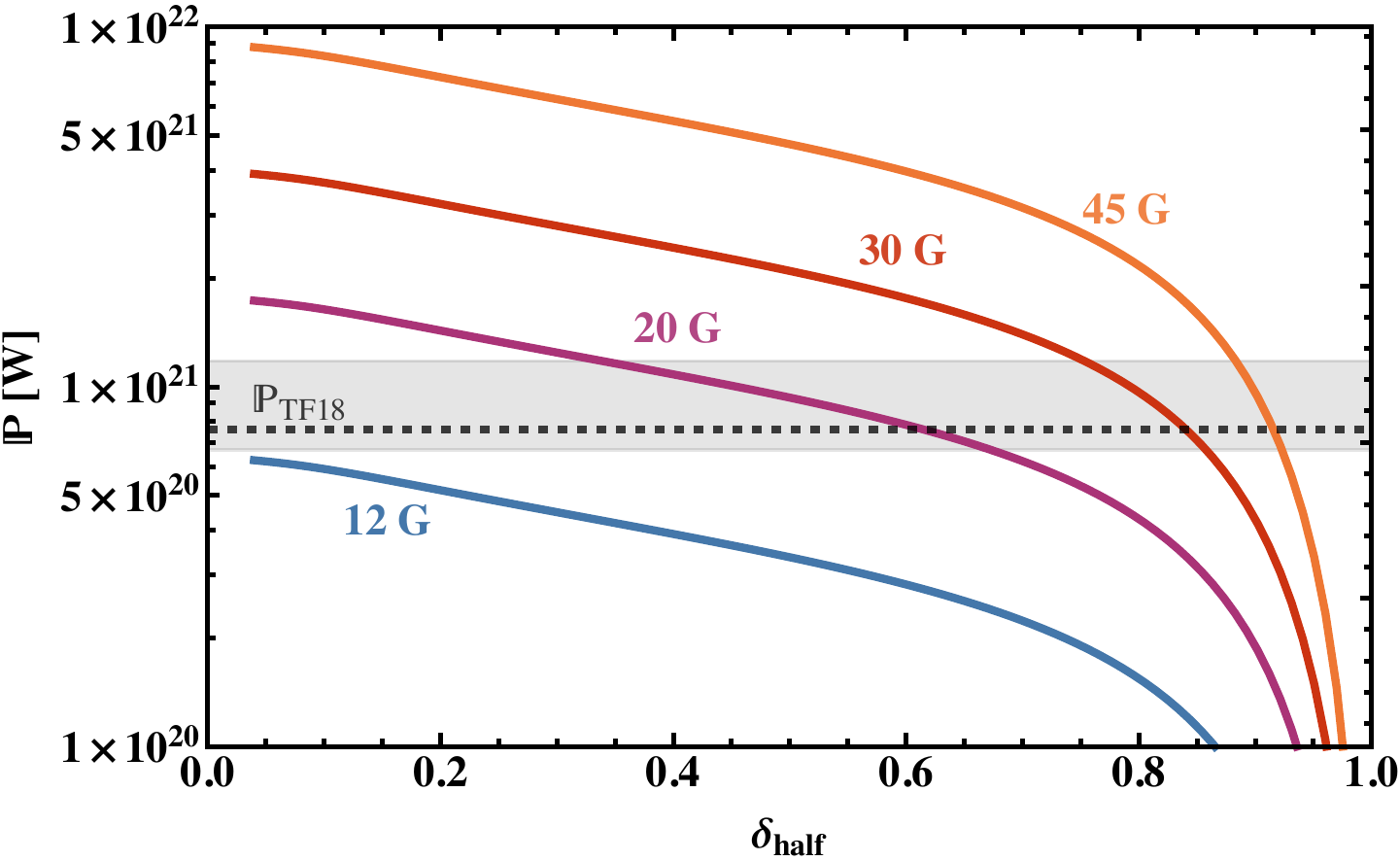}
        }
\caption{OD rate as a function of the wind profile $ \subs{\delta}{half}$ for HD 209458 b with $ \subs{v}{max}=\SI{2}{\km \per \s}$ \citep{Snellen2010}. The colors represent different magnetic field strengths at the equator, and the gray region indicates the anomalous heating rate computed by \cite{Thorngren_2018}.}
\label{fig:P_vs_delta_half}
\end{figure}

We computed $\mathds{P}$ in the range of $ 0 < \subs{\delta}{half} < 1$, which represents both shallow and deep zonal flows (Fig. \ref{fig:velocity}). The zonal wind profile influences the OD rate by an order of magnitude (Fig. \ref{fig:P_vs_delta_half}).
This is expected since the volume that contributes to the generation of the induced current (Eq. \ref{eq:j_ind}) increases. More precisely: the OD rate scales roughly like the volume integral, $\int \vec v \dint V$.

In addition, the anomalous heating rates of \citet{Thorngren_2018} are easily reproduced for magnetic fields stronger than $\sim \SI{12}{\gauss}$. Moreover, for HD 209458 b, we expect $ \subs{\delta}{half} = 0.82$ ($ \xi= \SI{51}{\per \Jupiterradius}$, $P(\delta_\mathrm{half})=\SI{0.34}{\bar}$) for $ \SI{27.72}{\gauss}$ \citep[scaling law from][]{Yadav2017} and $ \subs{v}{max} = \SI{2}{\km \per \s}$ \citep{Snellen2010}. As $P \propto e^{-r/H}$, $v \propto e^{\xi r} \propto P^{- \xi H} \approx P^{-0.52}$, which falls more slowly than the $v \propto P^{-5/2}$ found by \cite{Ginzburg_2016} and confines the circulation to the top $3\%$ of the HJ. In other words, our results indicate that a shallow weather layer is sufficient to explain the radius anomaly. 
%subsection: Zonal Wind Velocities
\subsection{Zonal wind velocities}
In the last section, we derived the flow geometry ($ \subs{\delta}{half}$) given the equatorial wind velocity. Conversely, we can ask what zonal wind velocity is required, given a flow geometry, to inflate a HJ sufficiently.

For this, we probed three HJ masses -- $\SI{0.5}{\Jupitermass}$, $\SI{1}{\Jupitermass}$, and $\SI{3}{\Jupitermass}$ -- for three zonal flow configurations: respectively, a linear decrease ($\subs{\delta}{half}=0.5$), a shallow weather layer similar to that of HD 209458 b ($\subs{\delta}{half}=0.8$), and a very shallow weather layer ($\subs{\delta}{half}=0.9$). The intrinsic temperature, HJ radius, and magnetic field strength were taken from the results of \cite{Thorngren_2018} and \cite{Yadav2017}.

\begin{figure}
\centering
\resizebox{\hsize}{!}{
                \includegraphics{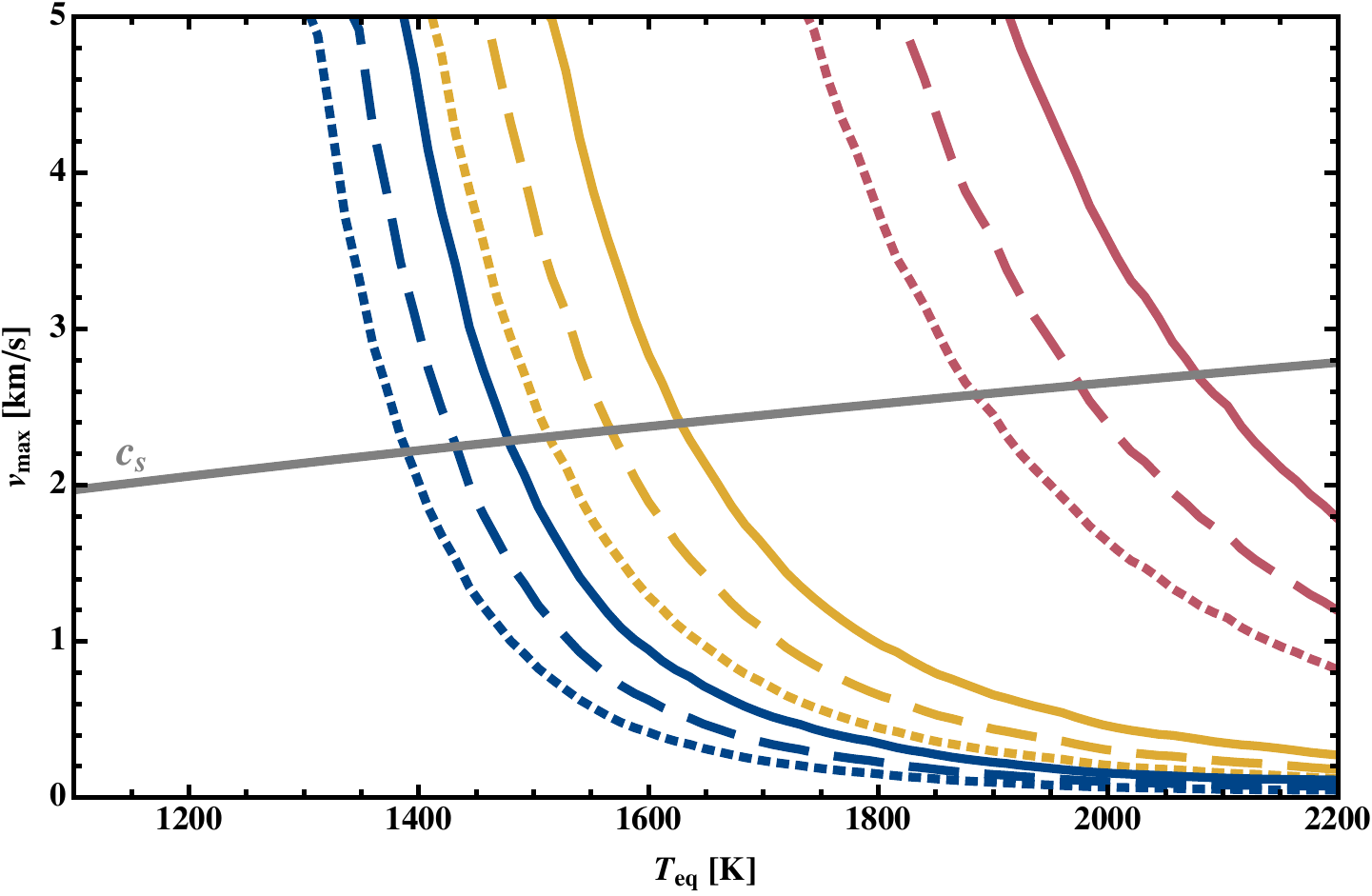}
        }
\caption{Equatorial wind velocity $ \subs{v}{max}$ needed to reproduce the anomalous heating powers of \citet{Thorngren_2018} as a function of equilibrium temperature. Blue represents $ \SI{0.5}{\Jupitermass}$, yellow $ \SI{1}{\Jupitermass}$, and red $ \SI{3}{\Jupitermass}$. The dotted, dashed, and solid lines represent $ \subs{\delta}{half}=0.5$, $0.8$, and $0.9$, respectively.}
\label{fig:v_match_TF17}
\end{figure}

The resulting velocities decrease with equilibrium temperature like a power law (Fig. \ref{fig:v_match_TF17}).
%
%Figure:v_match_TF17
%
This originates from a roughly linear increase in the anomalous heating power and the magnetic field with equilibrium temperature (i.e., $\mathds P \propto T_\mathrm{eq}$ and $B \propto T_\mathrm{eq}$). Since $\mathds P \propto v^2 B^2$, this leads to $v \propto \sqrt{\mathds{P}}/B \propto T^{-1/2}$. The physical interpretation for this effect is the damping of zonal flows due to strong magnetic drag.

Once we reach higher equilibrium temperatures, however, the magnetic field strength stays roughly constant \citep{Yadav2017}. Therefore, the velocity stays roughly constant as well. Furthermore, lower $ \subs{\delta}{half}$ values (i.e., more extended zonal jets) lead to lower velocities since OD is more efficient (see also the discussion below).

On the lower end of the spectrum, the equilibrium temperatures are too low to ionize sufficient amounts of alkali metals and planetary magnetic fields are weak. Hence, the zonal winds flow with little resistance, but also deposit little power into the planet's interior. Ultimately, the velocities diverge, indicating that OD cannot inflate efficiently at low $ \subs{T}{eq}$ -- an arguably trivial result. These extremely high velocities can be interpreted as upper bounds, which lose their physical meaning once the speed of sound is significantly exceeded and shock-induced dissipation sets in.
%
%Section: Discussion
%
\section{Discussion}\label{sec:discussion}
In this work, we have explored the physics of OD in HJs and derived an analytic expression for the power deposited into the convective interior as a function of wind penetration depth. This equation (see Eq. \ref{eq:P}) gives a simple and robust order-of-magnitude estimate.

We found that the wind profile has an order-of-magnitude influence on the power output. This effect must be taken into account in any detailed model of OD. Additionally, the scaling law allows unobservable planetary parameters to be constrained from direct observations. If, for example, one obtained the equatorial wind velocity of a HJ via spectroscopy, our scaling relation would predict the flow geometry with little effort.

Shallow weather layers appear sufficient to explain the radius anomaly. For HD 209458 b, we predict that the flow velocity drops by a factor of 2 at $\sim \SI{0.4}{\bar}$ compared to the wind velocity at the photosphere. While this velocity slope is rather steep, the Richardson number still exceeds 0.25, indicating that the Kelvin-Helmholtz instability never sets in. The weather layer size is on the lower end of GCM simulations and on the higher end of MHD simulations (Sect. \ref{sec:intro}). This model does not apply to low equilibrium temperatures ($\sim \SI{1000}{\kelvin}$), though, where a radius excess above the Jovian radius can be attributed to an extended radiative layer -- without the need for an interior heating mechanism.

\begin{figure}
\centering
\resizebox{\hsize}{!}{
                \includegraphics{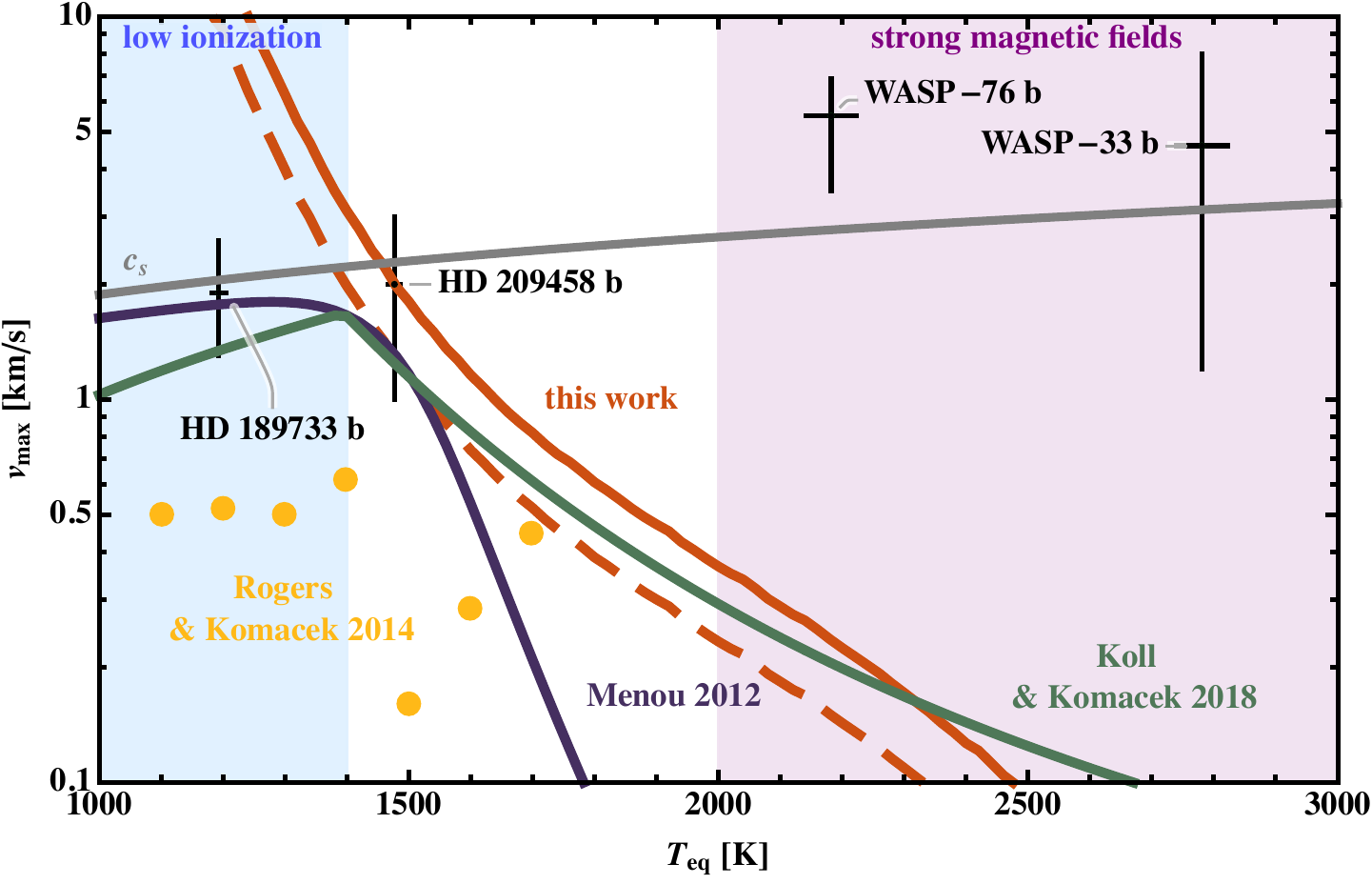}
        }
\caption{
Zonal wind velocity as a function of equilibrium temperature computed by \cite{Menou2012} (purple), \cite{Rogers_2014} (yellow), \cite{Koll_2018} (green), and this work (red). Our solid line indicates the equatorial wind velocity as computed by \cite{Menou2012} and our dashed line indicates the mean zonal wind velocity as computed by \cite{Rogers_2014} and \cite{Koll_2018}. The planetary parameters of HD 209458 b and a magnetic field of $ B = \SI{15}{\gauss}$ were held constant throughout our simulation. The velocity measurements are taken from \cite{Snellen2010} (HD 209458 b), \cite{Louden_2015} (HD 189733 b), \cite{Seidel:2021tm} (WASP-76 b), and \cite{Cauley_2021} (WASP-33 b).
}
\label{fig:v_model_comparison}
\end{figure}

Lastly, we found that, for a constant $\delta_\mathrm{half}$, the equatorial wind speed decreases with equilibrium temperature like a power law
\begin{align}
v_\mathrm{max}( \subs{T}{eq}) \propto (\subs{T}{eq}-T_0)^{-\alpha},
~\alpha\approx 1/2,
\end{align}
where $T_0$ is a constant that equals $T_0 \sim \SI{1000}{\kelvin}-\SI{1800}{\kelvin}$ depending on planet-specific parameters (radius, mass, etc.) --
before saturating at a few hundred meters per second. This profile shows a similar decrease to that in \cite{Menou2012}, \cite{Rogers_2014}, and \cite{Koll_2018} at around $\subs{T}{eq} \approx \SI{1500}{\kelvin}$  (Fig. \ref{fig:v_model_comparison}).
%
%Figure:v_model_comparison
%
In contrast, though, \cite{Menou2012} suggests more rapid velocity damping due to magnetic drag toward higher equilibrium temperatures.
Moreover,  \citet{Rogers_2014} arrive at considerably lower wind velocities (and hence OD rates) due to their numerical treatment ($\sim \SI{2e18}{\watt}$ for $ B= \SI{15}{\gauss}$ and $ v= \SI{160}{\m\per\s}$ at depth).
All of these models predict lower velocities than recently observed in ultra-hot Jupiters \citep[e.g., $\SI{5.5}{\km \per \s}$ for WASP-76 b,][]{Seidel:2021tm}. In this vein, our model suggests that only a very shallow circulation layer is required to explain the radius inflation in such a system, although a massive high-metallicity core could undoubtably alter this assertion. More fundamentally, our model assumptions are rendered invalid in the ultra-hot Jupiter regime ($\gtrsim \SI{2000}{\kelvin}$), where magnetic Reynolds numbers greatly exceed unity. Another distinction in atmospheric wind measurements between $T_\mathrm{eq}< \SI{2000}{\kelvin}$ HJs and ultra-hot Jupiters may lie in the fact that the sampled pressure levels of the atmosphere are vastly different. Ultimately, a more detailed analysis of MHD effects \citep[e.g.,][]{Rogers_2017,Hindle:2019um,beltz_2021} is required to understand their extremely fast wind velocities.

While recent theoretical evidence points toward enhanced bulk metallicities of HJs \citep[e.g.,][see also Appendix \ref{sec:metallicities}]{Thorngren2016, Mueller2020, Schneider2021}, those effects are subdominant to the uncertainty in the magnetic field and the velocity ($\mathds{P} \propto \sqrt{Z} v^2 B^2$). Nonetheless, since OD exclusively predicts a correlation between alkali metal abundance and radius inflation, atmospheric enrichment in alkali metals, as observed by \citet{Welbanks_2019}, might be an important contributing factor for this mechanism. For this study, though, as we do not claim to predict the exact anomalous heating power for {a specific planet} but rather introduce a general and robust extension to the Ohmic heating paradigm, they are of little concern for the validity of our results.

As the atmospheric composition, the wind velocity, and the dipole field strength of HJs are currently poorly constrained, we emphasize that this study can only yield an estimate of the broad circulation profile, rather than a precise prediction.\footnote{Temperature inversion in the atmosphere (i.e., in the presence of TiO) already alters the conductivity (and hence the OD rate) by two orders of magnitude.}
Likewise, more observational data of wind velocities in HJs, especially in the $T_\mathrm{eq}\sim \SI{1500}{\kelvin}$ range, would aid in constraining $ \subs{\delta}{half}$ empirically. Lastly, OD in the ultra-hot Jupiter regime remains poorly understood. Future studies could build upon recent analytical \citep{Hindle:2019um,Hindle_2021} or numerical \citep{beltz_2021} work to account for those objects self-consistently. Nonetheless, this work shows that OD cannot only explain the radius inflation problem, but also allows us to peak deep into HJ atmospheres and, thereby, gives us new insights into their remarkable structure.
\begin{acknowledgements}
B.B., acknowledges the support of the European Research Council (ERC Starting Grant 757448-PAMDORA).  K.B. is grateful to Caltech, and the David and Lucile Packard Foundation for their generous support.
\end{acknowledgements}

\bibliographystyle{aa}
\bibliography{42588corr}
\begin{appendix}
\begin{onecolumn}
\section{Solving the electric potential equation} \label{sec:solving_phi}
The conductivity profile in this work is given by
\begin{align}
\sigma (r) = 
\begin{cases}
\sigma_\mathrm{RCB} \exp\left(\frac{r-r_\mathrm{RCB}}{2 H}\right) & \mathrm{for}~r_\mathrm{RCB} < r \leq R\\
\sigma_\mathrm{RCB} \exp\left(\frac{r_\mathrm{RCB}-r}{H_\mathrm{fit}}\right) & \mathrm{for}~r_\mathrm{max}<r \leq r_\mathrm{RCB}\\
\sigma_\mathrm{max} & \mathrm{for}~r\leq r_\mathrm{max}\\
\end{cases},
\end{align}
where $\subs{r}{max}$ is the radius at which the conductivity reaches the maximum value of $\sigma_\mathrm{max} = \SI{1e6}{\siemens \per \m}$.
Accordingly, we solved Eq. \ref{eq:electric_potential} in those three regions, assuming a dipole magnetic field
\begin{align}
\vec B =\nabla \times k_m \left(\frac{\sin \theta}{r^2}\right)\hat\phi,
\end{align}
where $k_m$ is the dipole magnetic moment.

This was done by decomposing the equation into spherical harmonics $Y_l^m (\theta, \phi)$, which yields that only the $l=2$ and $m=0$ component contribute to the OD rate. We obtained the constants of integration by enforcing continuity, assuming that the electric potential is zero at the center, $\Phi (r=0) = 0$, and that there is an insulating boundary layer at $R$: $\vec j_\mathrm{ind} (R,\theta,\phi) \cdot \hat r = 0$. Typical values for $R$ include the transit radius of the planet or the radius at which the conductivity in the radiative layer drops significantly. The resulting induced current density is evaluated as described in Sect. \ref{sec:OD_theory}. The OD rate becomes
\begin{align}\label{eq:P}
\mathds P = v_\mathrm{max}^2 B^2 \sigma_\mathrm{RCB} \cdot f(r_\mathrm{RCB},R, R_\mathrm{p}, H, H_\mathrm{fit},\xi),
\end{align}
where $f$ is given by
\begin{align}
\begin{split}
&- 
64 \pi  H^2 H_{\text{fit}} R_\mathrm{p}^6 \left(4
H_{\text{fit}}-r_{\text{RCB}}\right) \left(-6
   H_{\text{fit}} r_{\text{RCB}}+12
   H_{\text{fit}}^2+r_{\text{RCB}}^2\right)
   \\
   &\cdot
   \Bigg(\xi
   ^2 (2 H \xi +1) (6 H \xi +1) r_{\text{RCB}}^3
   \bigg(e^{\frac{R}{2 H}} \left(48 H^2-12
   H R+R^2\right)
   \left(\text{Ei}\left(\xi 
   R\right)-\text{Ei}\left(\xi 
   r_{\text{RCB}}\right)\right)
   \\
   &-\left(48 H^2+12 H
   R+R^2\right)
   \left(\text{Ei}\left(\xi 
   R+\frac{R}{2
   H}\right)-\text{Ei}\left(\xi 
   r_{\text{RCB}}+\frac{r_{\text{RCB}}}{2
   H}\right)\right)\bigg)
   \\
   &+r_{\text{RCB}}^3
   \left(-e^{\frac{R}{2 H}+\xi 
   R}\right) \left(\xi  \left(-24 H^2 \xi 
   (12 H \xi +5)+6 H \xi 
   R+R\right)-1\right)
   \\
   &+e^{\xi 
   r_{\text{RCB}}} \bigg(e^{\frac{R}{2 H}}
   \Big(24 H^2 \xi  \left(48 H^2-12 H
   R+R^2\right)+\xi  (2 H \xi
   +1) (6 H \xi +1) r_{\text{RCB}}^2 \left(48 H^2-12 H
   R+R^2\right)
   \\
   &+4 H \xi  (3 H
   \xi +2) r_{\text{RCB}} \left(48 H^2-12 H
   R+R^2\right)-r_{\text{RCB}}^
   3\Big)
   \\
   &-2 H \xi  \left(48 H^2+12 H
   R+R^2\right)
   e^{\frac{r_{\text{RCB}}}{2 H}} \left(6 H \left(\xi 
   r_{\text{RCB}} \left(\xi 
   r_{\text{RCB}}+1\right)+2\right)+r_{\text{RCB}}
   \left(\xi 
   r_{\text{RCB}}-2\right)\right)\bigg)\Bigg)^2
   \\
   &/
   \Bigg \{5
   r_{\text{RCB}}^3 \left(e^{\xi  R}-e^{\xi
    r_{\text{RCB}}}\right)^2 \bigg(\left(48 H^2+12 H
   R+R^2\right)
   e^{\frac{r_{\text{RCB}}}{2 H}}
   \\
    &\cdot\left(288 H^2
   H_{\text{fit}} \left(H_{\text{fit}}+2 H\right)-6
   \left(H_{\text{fit}}+2 H\right) r_{\text{RCB}}^3+12
   \left(H_{\text{fit}}+2 H\right)
   \left(H_{\text{fit}}+3 H\right) r_{\text{RCB}}^2-96
   H \left(H_{\text{fit}}+H\right)
   \left(H_{\text{fit}}+2 H\right)
   r_{\text{RCB}}+r_{\text{RCB}}^4\right)
   \\
   &-12 H
   \left(H_{\text{fit}}+2 H\right)
   e^{\frac{R}{2 H}} \left(48 H^2-12 H
   R+R^2\right) \left(4
   H_{\text{fit}} r_{\text{RCB}}+24 H H_{\text{fit}}-8
   H r_{\text{RCB}}-r_{\text{RCB}}^2\right)\bigg)^2 \Bigg\},
\end{split}
\end{align}
with $\text{Ei}(z) := -\int_{-z}^{\infty} e^{-t}/t~\dint t$.
\end{onecolumn}
\begin{twocolumn}
\section{Influence of metallicities} \label{sec:metallicities}
The Solar System's giant planets are enhanced in metals compared to solar values \citep[e.g.,][]{Atreya2016, Li2020}. While it is not yet clear how planets form, interior modeling \citep[e.g.,][]{Thorngren2016, Mueller2020} and planet formation models \citep[e.g.,][]{Schneider2021} suggest that similarly enhanced metallicities are plausible for HJs. 

The influence of metals on the OD rate is a trade-off between increased electric conductivity and atmospheric contraction due to higher mean molecular weight. For low metallicities, the increase in conductivity dominates, leading to $\mathds{P} \propto \sqrt Z$. Increasing the metallicity further, contraction starts to play a significant role, damping the dissipation rate (Fig. \ref{fig:metallicity}). Nonetheless, the change in $\mathds{P}$ is clearly subdominant to the uncertainty in velocity and magnetic field strength.
\begin{figure}
\centering
\resizebox{\hsize}{!}{
                \includegraphics{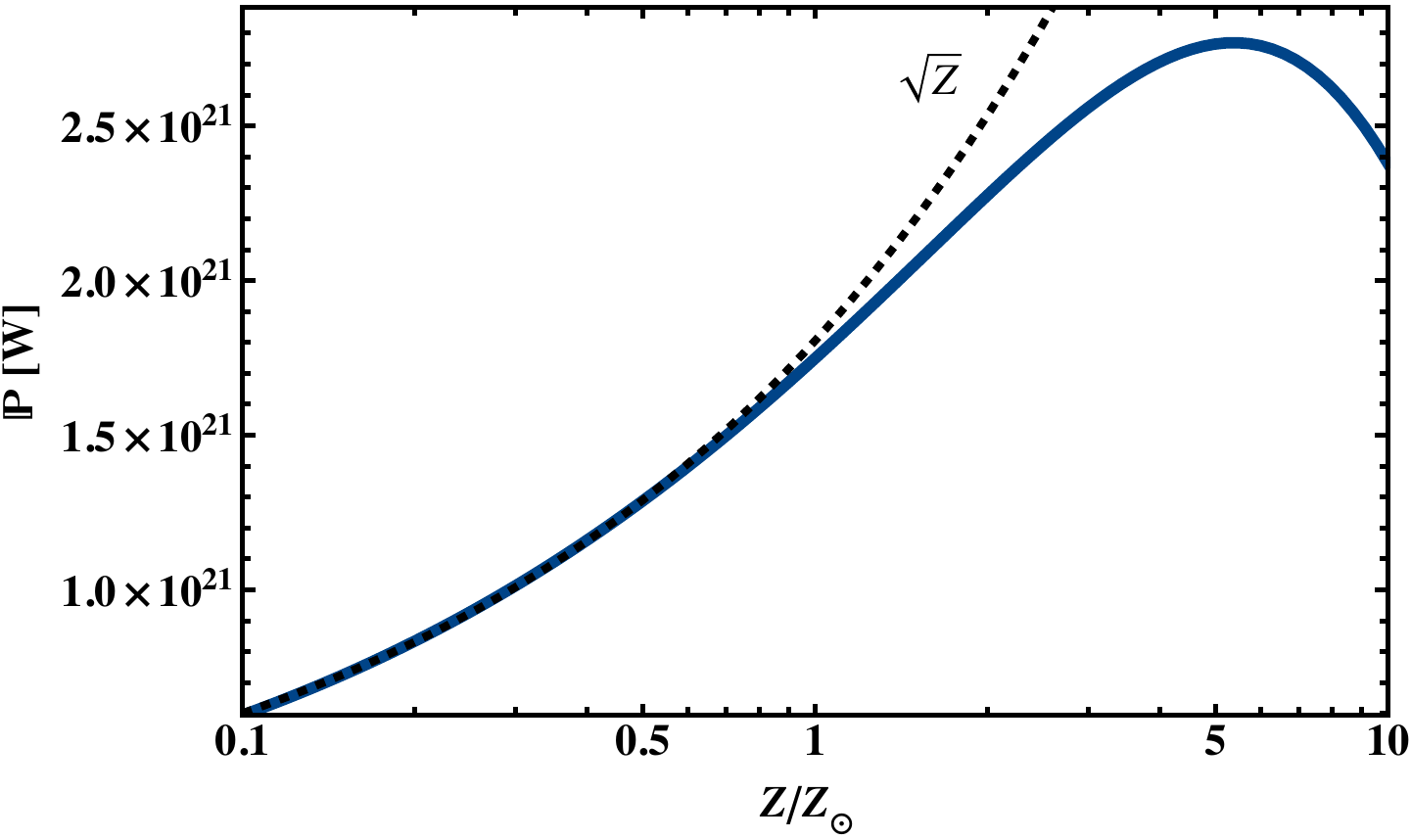}
        }
\caption{OD rate as a function of metallicity assuming the parameters of HD 209458 b and a linearly decreasing jet ($\delta_\mathrm{half}=0.5$).}
\label{fig:metallicity}
\end{figure}
%
%%%%%%%%%%%%%% RCB
%
\section{Shallow RCB layers} \label{sec:RCB_depth}
As mentioned in Sect. \ref{sec:theory}, the depth of the RCB differs significantly between studies that use the Schwarzschild criterion \citep[e.g.,][]{Burrows:2003wc, Batygin2011, Kumar_2021} and recent statistically motivated studies \citep{Thorngren2019, Sarkis:2021tr}. The latter arrive at considerably shallower RCB depths, which may be a consequence of depositing the entire heat in the center of the planet. In fact, \cite{Youdin_2010} argue that turbulent diffusion in the radiative layer pushes the RCB down to the kilobar range. In any case, the penetration depth depends only weakly on the exact value (Fig. \ref{fig:half_velocity_pressure}).
\begin{figure}
\centering
\resizebox{\hsize}{!}{
                \includegraphics{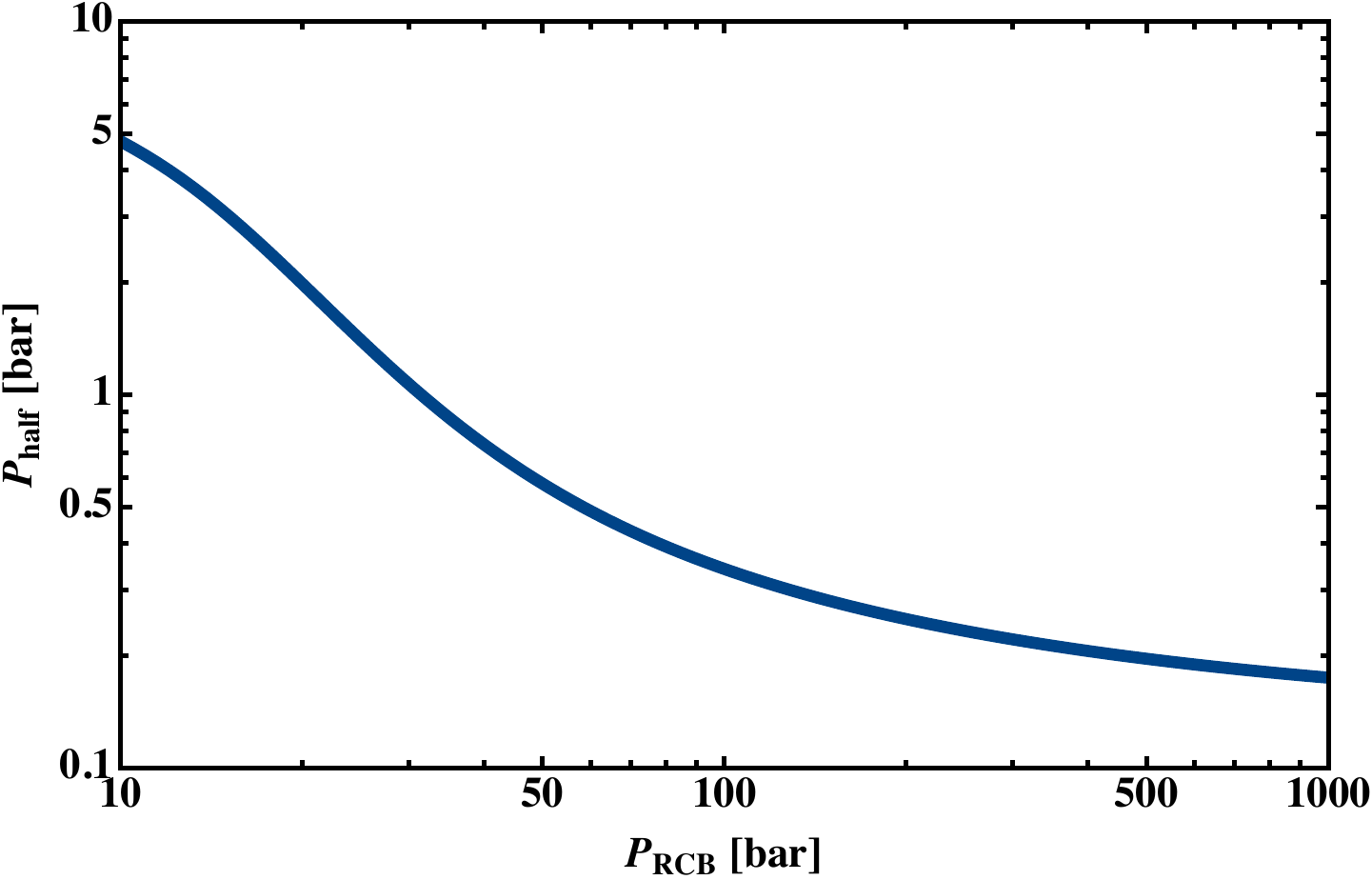}
        }
\caption{Half velocity pressure, $P_\mathrm{half}$, as a function of RCB pressure, $P_\mathrm{RCB}$, assuming the parameters of HD 209458 b.}
\label{fig:half_velocity_pressure}
\end{figure}
Hence, we conclude that our results are robust against variations in the RCB depth.
\end{twocolumn}
\end{appendix}
\end{document}